\documentclass[a4paper,11pt]{article}
\pdfoutput=1 

\usepackage{jheppub1} 

\usepackage[T1]{fontenc} 
\usepackage{empheq,mathtools}
\usepackage{cancel,slashed}
\usepackage{xcolor}

\def\tsp{\hspace{0.083333em}}
\def\arXiv#1{\href{http://arxiv.org/abs/#1}{arXiv:#1}}
\def\arXiv#1#2{\href{http://arxiv.org/abs/#1}{arXiv:#1}}
\def\arXivid#1#2{\href{http://arxiv.org/abs/#1/#2}{#1/#2}}
\def\doi#1#2{\href{http://doi.org/#1}{#2}}

\title{\boldmath Holographic superconductors at zero density}

\author{Jie Ren and Haodong Xie}

\affiliation{School of Physics, Sun Yat-sen University, Guangzhou, 510275, China}

\emailAdd{renjie7@mail.sysu.edu.cn}

\abstract{We construct holographic superconductors at zero density. The model enjoys a luxury property that the background geometry dual to the ground state is analytically available. It has a hyperscaling-violating geometry in the IR and is asymptotically AdS in the UV. Classification by IR geometries gives new insights on supergravity solutions. We numerically construct the finite temperature solution of hairy black holes and verify the phase transition by tuning a double-trace deformation parameter. For a holographic superconductor from M-theory, we obtain an analytic solution of the AC conductivity, which explicitly shows a superconducting delta function and a hard gap.}

\arxivnumber{2206.03498}

\begin{document}
\maketitle
\flushbottom

\section{Introduction}
\label{sec:intro}
Holographic superconductors are anti-de Sitter (AdS) black holes that can spontaneously develop a hair \cite{Hartnoll:2008vx,Hartnoll:2008kx}, and they are successful applications of the AdS/CFT correspondence \cite{Maldacena:1997re,Gubser:1998bc,Witten:1998qj} to the study of quantum matter without quasiparticles \cite{Hartnoll:2016apf}. Strongly interacting quantum field theories are studied in terms of tractable models of classical gravity, including consistent truncations of ten- and eleven-dimensional supergravities. The ``minimal'' model of holographic superconductors has a charged black hole and a complex scalar field. The black hole has instability, and develops a scalar hair below a critical temperature. This bulk scalar field is dual to the order parameter of the spontaneous breaking of the U(1) symmetry. There are two types of instability. One is triggered by a zero mode, which is a pole of the retarded Green's function at the frequency $\omega=0$ \cite{Gubser:2008px}. The other is the IR instability, which happens when the exponent of the near-horizon AdS$_2$ geometry becomes imaginary \cite{Hartnoll:2008kx,Faulkner:2009wj}. Both types of instability can happen when the charge of the scalar field is sufficiently large.

Although holographic superconductors were intensely studied during the past 14 years (see \cite{Cai:2015cya} for a review), exact, analytic solutions of either the background geometry dual to the superconducting phase or the frequency-dependent conductivity are extremely rare. We will show that holographic superconductors at zero density enjoy much more analytic controls. While charged black holes are dual to finite density systems, neutral black holes are dual to zero density systems, which also play an important role in condensed matter physics. Zero density systems are particle-hole symmetric, and have an electrical conductivity due to pair production. Recently, an unprecedentedly large number of analytic solutions of the AC conductivity have been obtained \cite{Ren:2021rhx}.

Zero density systems can be superconducting, and can be realized by the frustrated Hubbard model at half filling on a square lattice \cite{Hubbard}
\begin{equation}
H=t\sum_{\langle i, j\rangle}c_i^\dagger c_j+t'\sum_{\langle\langle i, j\rangle\rangle}c_i^\dagger c_j+U\sum_i n_{i\uparrow}n_{i\downarrow}\,,
\end{equation}
where $t$ is the nearest neighbor hopping, $t'$ is the next-nearest neighbor hopping, and $U$ is the interaction. It was shown that superconductivity occurs in the phase diagram as a function of $U/t$ and $t'/t$.  It is desirable to construct a holographic dual to a superconductor at zero density as a simplified version of holographic superconductors, by retaining essential properties of superconductivity while having more analytic controls.

A mechanism for holographic superconductors at zero density was proposed in \cite{Faulkner:2010gj}. The superconducting instability is triggered by a zero mode when a multi-trace deformation is present, and the neutral black hole develops a scalar hair below a critical temperature. The system shares the same background geometry as an Einstein-scalar system, while the gauge field plays a role for fluctuations. Another mechanism for holographic superconductors at zero density involves more interactions \cite{Basu:2019pxw}, and we consider the first one here. A key advantage of our model is that the ground state and the corresponding AC conductivity are analytically solvable.

The ground state of our model is obtained by taking a nontrivial neutral limit of analytic solutions in an Einstein-Maxwell-dilaton (EMD) system whose special cases intersect with STU supergravities. It was shown that there are two neutral limits for this class of charged dilaton black hole solutions \cite{Ren:2019lgw}; see appendix~\ref{sec:GR} for a special example. In the trivial neutral limit, we obtain the Schwarzschild-AdS black hole. In the nontrivial neutral limit, we can consistently turn off the gauge field while keeping the dilaton field nontrivial. As a consequence, we obtain analytic solutions of an Einstein-scalar system with a spacetime singularity in the IR.\footnote{We study planar black holes throughout this paper. Only in the hyperbolic case does this nontrivial neutral limit give a black hole \cite{Ren:2019lgw}.} The IR is a hyperscaling-violating geometry, and the UV is asymptotically AdS. In this paper, we employ this geometry as the ground state of holographic superconductors at zero density.

It is common to have a naked singularity in the IR with a running scalar field. The solution is acceptable if the singularity can be resolved, for example, many supergravity solutions with a singularity can be lifted to ten or eleven dimensions without singularities. In \cite{Gubser:2000nd}, Gubser proposed a criterion to justify naked singularities. There are two statements:
\begin{itemize}
\item[(A)] The scalar potential is bounded from above in the solution.
\item[(B)] The geometry can be obtained as the extremal limit of a regular black hole.
\end{itemize}
Statement~(B) is a weak form of cosmic censorship and implies statement~(A) \cite{Gubser:2000nd}. Although it may be too strong, statement~(B) has broader applications, and is often called the Gubser criterion in the literature. The two statements agree in our model.

If the Gubser criterion is satisfied, the ground state of our model can be obtained by taking the extremal limit of a finite temperature black hole. Unfortunately, the finite temperature solution is not analytically available. By choosing a typical value of the parameter in our model~\eqref{eq:action} below with the scalar potential~\eqref{eq:3-exp-V} and $A_\mu=0$, we numerically construct the finite temperature solution with a sourceless boundary condition given by a double-trace deformation in the dual CFT. The temperature changes by tuning the parameter for the double-trace deformation. In other words, the analytic solution for the ground state lacks one parameter, which can be numerically added as the temperature. Scalar condensation happens below a critical temperature. We carefully compare the free energy among different saddles and show that the dominant one has the ground state with a hyperscaling-violating geometry in the IR.

A summary of our results is as follows:

\begin{itemize}
\item Starting with the analytic solution of the ground state, we analyze its IR geometry by classifying the gapless and gapped states. Special cases belonging to STU supergravity are identified in the parameter space at the separation of gapless and gapped states.

\item We numerically construct the finite temperature solution and show that there are phase transitions. We calculate the free energy, and show that the hairy black hole has lower free energy lower under $T_c$.

\item For studying the superconducting phase, we calculate the AC conductivity, and verify that there is a superconducting delta function, and a sum rule is satisfied as the temperature varies.

\item For a holographic superconductor from M-theory, we analytically solve the AC conductivity, which unambiguously shows a delta function and a hard gap.

\end{itemize}

In section~\ref{sec:ES}, we review the key ingredients of holographic superconductors at zero density. In section~\ref{sec:GS}, we analyze the ground state solutions, and classify them by their IR geometries. In section~\ref{sec:SC}, we numerically construct the finite temperature solution and calculate the conductivity. In section~\ref{sec:MT}, we obtain an analytic solution of the AC conductivity for a holographic superconductor from M-theory. In section~\ref{sec:general}, we present a more general solution and discuss the relation between finite density and zero density systems. In section~\ref{sec:sum}, we conclude. In appendix~\ref{sec:HR}, we give more details on the holographic renormalization. In appendix~\ref{sec:GR}, we comment on the neutral limit of the Gubser-Rocha model. In appendix~\ref{sec:KU}, we show that for the near-extremal solution, the geometry near the spacetime singularity is a Kasner universe. In appendix~\ref{sec:AdS5}, we present the AdS$_5$ solutions.

\section{From Einstein-scalar systems to holographic superconductors}
\label{sec:ES}
We start with a general model of St\"{u}ckelberg holographic superconductors \cite{Franco:2009yz}, which have spontaneous breaking of the U(1) symmetry at low temperatures. The Lagrangian density is\footnote{The original model of holographic superconductors \cite{Hartnoll:2008kx} with $\psi=\phi e^{ip}$ is a special case of \eqref{eq:action}:
\begin{equation}
\mathcal{L}=R+\frac{6}{L^2}-\frac{1}{4}F_{\mu\nu}F^{\mu\nu}-|\nabla\psi-iqA\psi|^2-V(|\psi|)\,.\nonumber
\end{equation}}
\begin{equation}
\mathcal{L}=R-{1\over 2}(\partial\phi)^2-V(\phi)-{Z(\phi)\over 4}F_{\mu\nu}F^{\mu\nu}-\frac{\mathcal{W}(\phi)}{2}(\partial_\mu p-A_\mu)^2,\label{eq:action}
\end{equation}
where $F=dA$, and the functions have the form
\begin{equation}
V=-\frac{6}{L^2}+\frac{1}{2}m^2\phi^2+\mathcal{O}(\phi^3),\qquad
Z=1+g_Z\phi^2+\mathcal{O}(\phi^3),\qquad
\mathcal{W}=g_\mathcal{W}\phi^2+\mathcal{O}(\phi^3)\,.
\end{equation}
The gauge symmetry is $A_\mu\to A_\mu+\partial_\mu\alpha$ and $p\to p+\alpha$. We choose the gauge $p=0$. In the original model of holographic superconductors, we can think of $\phi$ and $p$ as the magnitude and phase of a complex scalar field. Nevertheless, in the more general St\"{u}ckelberg holographic superconductors \cite{Franco:2009yz}, $\phi$ and $p$ are any real scalar fields, and $V(\phi)$, $Z(\phi)$, and $\mathcal{W}(\phi)$ are not necessarily even functions of $\phi$. We mainly study AdS$_4$ systems in this paper, and the generalization to higher dimensions is straightforward. As an example, the special choice of functions is from a consistent truncation of M-theory \cite{Donos:2011ut}:
\begin{equation}
V(\phi)=-\frac{2}{L^2}(\cosh\phi+2),\qquad Z(\phi)=1,\qquad \mathcal{W}(\phi)=\frac{1}{L^2}\sinh^2\Bigl(\frac{\phi}{2}\Bigr).
\end{equation}
We calculate the AC conductivity for the ground state of this model in section~\ref{sec:MT}.

We are interested in neutral black hole solutions to the system. When $A_\mu=0$, the above system shares the same background geometry as an Einstein-scalar system
\begin{equation}
\mathcal{L}=R-{1\over 2}(\partial\phi)^2-V(\phi)\,.\label{eq:ES}
\end{equation}
In other words, an Einstein-scalar system can be promoted to a holographic superconductor at zero density, if the scalar hair is spontaneously developed. In the following, we briefly review the mechanism of the scalar condensation achieved by multi-trace deformations \cite{Faulkner:2010gj}. When we study the fluctuation of the gauge field, such as calculating the conductivity, we need to use the action~\eqref{eq:action}.

The asymptotic behavior of the scalar field near the AdS boundary is
\begin{equation}
\phi =\phi_a \tilde{z}^{\Delta_-}(1+\cdots)+\phi_b \tilde{z}^{\Delta_+}(1+\cdots),\label{eq:phibry}
\end{equation}
where $\tilde{z}$ is the Fefferman-Graham (FG) radial coordinate,\footnote{Other coordinates may give different values of $\phi_a$ and $\phi_b$. An example is the solution~\eqref{eq:sol4}--\eqref{eq:fsol4} below, in which the near boundary expansion does not satisfy $\sqrt{U}=r(1+0\cdot r^{-1}+\cdots)$.}
and $\Delta_\pm=3/2\pm\sqrt{9/4+m^2L^2}$. In the standard quantization, $\phi_a$ is the source of the scalar operator $\mathcal{O}$ dual to $\phi$, and $\phi_b$ is its expectation value. When $-9/4<m^2L^2<-5/4$, both these modes are normalizable. We can modify the action by a double-trace deformation
\begin{equation}
S\to S-\kappa\int d^3x\,\mathcal{O}^2\,.
\end{equation}
To have a relevant deformation, we must start with the alternative quantization, in which the scalar operator has a zero source and expectation value $\langle\mathcal{O}\rangle=\phi_a$.
The double-trace deformation corresponds to a new boundary condition on $\phi$ \cite{Witten:2001ua,Berkooz:2002ug}:
\begin{equation}
\phi_b=\kappa\phi_a\,.
\label{eq:kappa}
\end{equation}
The new Green's function for the scalar operator is
\begin{equation}
G^{(\kappa)}=\frac{1}{G^{-1}+\kappa}\,,
\end{equation}
where $G=-\phi_a/\phi_b$ is the Green's function for the alternative quantization.

At a critical temperature $T_c$, the black hole is unstable against perturbations of the scalar field, which corresponds to a zero mode in the Green's function at $\phi_b=\kappa\phi_a$. This happens at a negative $\kappa$ \cite{Faulkner:2010gj}. The linearized equation for the scalar field $\phi$ is the Klein-Gordon equation $(\nabla^2-m^2)\phi=0$, which can be solved in terms of hypergeometric functions. We will use the solution with $m^2L^2=-2$. With the regularity boundary condition at the horizon, and \eqref{eq:kappa} at the AdS boundary, the critical temperature can be analytically obtained \cite{Faulkner:2010gj,Mefford:2014gia}:
\begin{equation}
\frac{T_c}{-\kappa}=\frac{3}{4\pi}\frac{\Gamma(4/3)\Gamma(1/3)^2}{\Gamma(2/3)^3}\approx 0.616\,.
\end{equation}
The double-trace deformation parameter $\kappa$ introduces another scale, and we use it as the unit in writing dimensionless quantities. Below the critical temperature, the black hole will develop a scalar hair.

To further support the phase transition, we will show that below $T_c$, the hairy black hole has lower free energy than the Schwarzschild-AdS black hole. If there is more than one hairy black hole solution (saddle), we take the one with the lowest free energy.

\section{Analytic solution for the ground state}
\label{sec:GS}

For future convenience, we define some terminologies. There are four U(1) gauge fields in STU supergravity in AdS$_4$ with charges $Q_{i}$. We call the system 3-charge black hole if $Q_1=Q_2=Q_3 =Q$ and $Q_4=0$; 2-charge black hole if $Q_1=Q_2 =Q$ and $Q_3=Q_4=0$; 1-charge black hole if $Q_1 =Q$ and $Q_2=Q_3=Q_4=0$. There are three U(1) gauge fields in STU supergravity in AdS$_5$ with charges $Q_{i}$. We call the system 2-charge black hole if $Q_1=Q_2 =Q$ and $Q_3=0$; 1-charge black hole if $Q_1 =Q$ and $Q_2=Q_3=0$. When all the charges are the same, the solution is the Reissner-Nordstr\"{o}m-AdS black hole. These black holes are asymptotically AdS solutions to Einstein-Maxwell-dilaton systems. They are the simplest and most reliable solutions in the applications of the AdS/CFT correspondence, as they can be embedded in ten or eleven dimensions \cite{Cvetic:1999xp}. We consider AdS$_4$ solutions in the following, and put AdS$_5$ solutions in appendix~\ref{sec:AdS5}.

Interestingly, there exists a one-parameter family of the scalar potential that interpolates among supergravity systems. The potential of the scalar field is given by \cite{Gao:2004tu}\footnote{This potential has been rediscovered by different people in different ways. One way is by adding a cosmological constant for the Garfinkle-Horowitz-Strominger (GHS) black hole \cite{Garfinkle:1990qj}.}
\begin{equation}
V(\phi)=-\frac{2}{(1+\alpha^2)^2L^2}\left[\alpha^2(3\alpha^2-1)e^{-\phi/\alpha}+8\alpha^2e^{(\alpha-1/\alpha)\phi/2}+(3-\alpha^2)e^{\alpha\phi}\right],\label{eq:3-exp-V}
\end{equation}
where $\alpha$ is a parameter, and the values of $\alpha=0$, $1/\sqrt{3}$, $1$, and $\sqrt{3}$ correspond to special cases of STU supergravity. The potential is invariant under $\alpha\to -\alpha$ and $\phi\to -\phi$, and we assume $\alpha>0$ without loss of generality. The values of $\alpha=1/\sqrt{3}$, $1$, $\sqrt{3}$ correspond to 3-charge, 2-charge, and 1-charge black holes in AdS$_4$, respectively. Only in these cases does the potential have a $\mathbb{Z}_2$ symmetry. The functions $Z(\phi)$ and $\mathcal{W}(\phi)$ in the action~\eqref{eq:action} do not affect the background geometry as a neutral black hole. We will specify them as \eqref{eq:ZW} below when we calculate the conductivity.

This potential contains the following information: (i) the cosmological constant, (ii) $V'(0)=0$, (iii) scaling dimension of the scalar operator in the UV, (iv) hyperscaling violation exponent in the IR. The $\phi\to 0$ behavior is $V(\phi)=-6/L^2-(1/L^2)\phi^2+\cdots$, where the first term is the cosmological constant, and the second term gives the mass of the scalar field $m^2L^2=-2$. The scaling dimension of the dual scalar operator in the CFT is $\Delta_-=1$ or $\Delta_+=2$. We start with the alternative quantization, and thus take the $\Delta_-=1$ operator.

\begin{figure}
\centering
\includegraphics[width=0.325\textwidth]{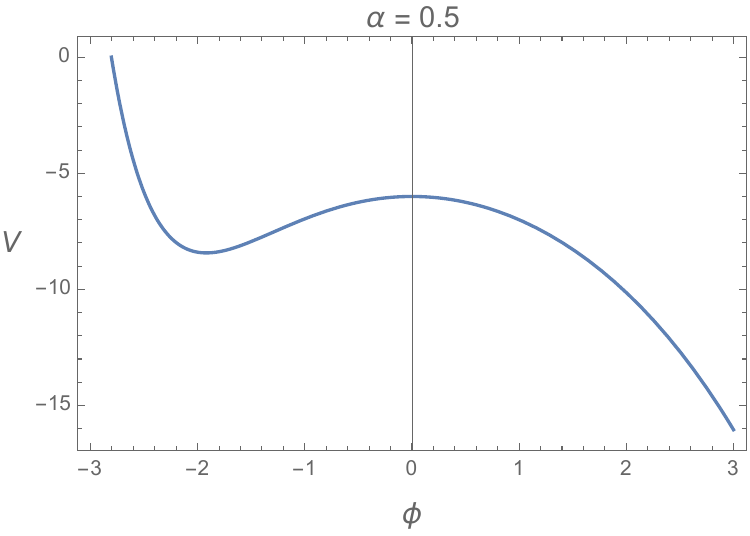}
\includegraphics[width=0.325\textwidth]{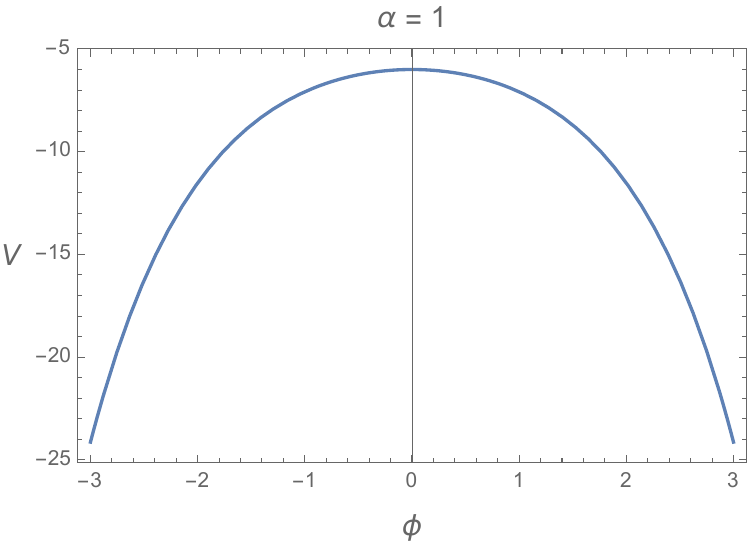}
\includegraphics[width=0.325\textwidth]{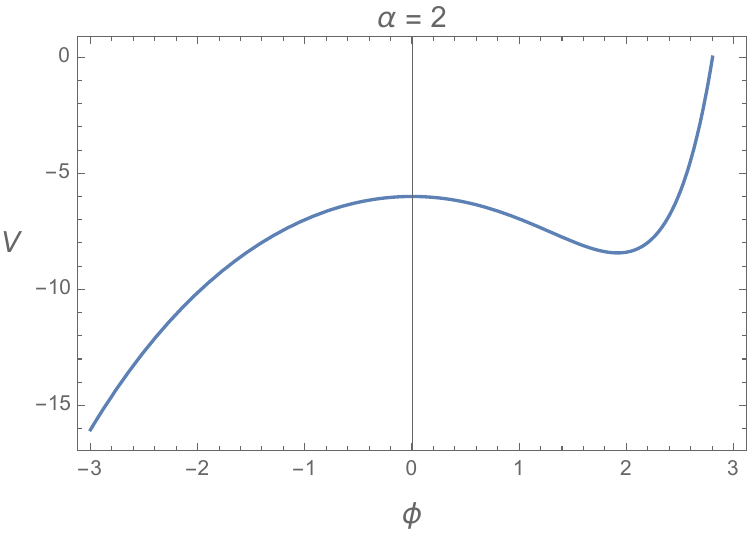}
\caption{\label{fig:Vphi} Shape of the potential $V(\phi)$. Note that this is different from the on-shell potential $V[\phi(r)]$ as a function of $r$ for the Gubser criterion.}
\end{figure}

The shape of the potential $V(\phi)$ is plotted in figure~\ref{fig:Vphi}, indicating possible IR geometries for the ground state. When $0<\alpha<1/\sqrt{3}$ or $\alpha>\sqrt{3}$, there is a local minimum of the potential at $V'(\phi_*)=0$, where
\begin{equation}
\phi_*=\frac{2\alpha}{1+\alpha^2}\log\biggl(\frac{1-3\alpha^2}{3-\alpha^3}\biggr),
\end{equation}
implying that there is a metastable solution whose IR is AdS$_4$. Another solution has a runaway scalar field in the IR. When there exist two black hole solutions with scalar hair, we can use the relation between the entropy and temperature to distinguish them.

The system~\eqref{eq:action} with the scalar potential~\eqref{eq:3-exp-V} admits a neutral solution with a nontrivial dilaton profile. The solution of the metric $g_{\mu\nu}$, gauge field $A_\mu$, and dilaton field $\phi$ is \cite{Ren:2019lgw}
\begin{align}
ds^2 &=f\tsp (-dt^2+d\vec{x}^2)+f^{-1}dr^2,\qquad A=0\,,\nonumber\\
f &=\frac{r^2}{L^2}\left(1-\frac{b}{r}\right)^\frac{2\alpha^2}{1+\alpha^2},
\qquad e^{\alpha\phi}=\left(1-\frac{b}{r}\right)^\frac{2\alpha^2}{1+\alpha^2}.
\label{eq:solution}
\end{align}
This is the nontrivial neutral limit of \eqref{eq:sol4}--\eqref{eq:fsol4} below with $a=c=\beta=0$. For $\alpha\neq 0$, this metric does not have a regular horizon, and has a spacetime singularity at $r=b$ if $b>0$, and at $r=0$ if $b<0$. We expect that this solution can be taken as an extremal limit of a finite temperature black hole solution, when the Gubser criterion \cite{Gubser:2000nd} is satisfied.

The IR geometry is a hyperscaling-violating geometry:
\begin{equation}
ds^2=\tilde{r}^\theta\left(-\frac{dt^2}{\tilde{r}^{2\mathsf{z}}}+\frac{d\tilde{r}^2+dx^2+dy^2}{\tilde{r}^2}\right),\label{eq:ztheta1}
\end{equation}
where the Lifshitz scaling exponent $\mathsf{z}$ and the hyperscaling violation exponent $\theta$ are \cite{Ren:2021rhx}
\begin{equation}
\mathsf{z}=1,\qquad \theta=\begin{cases}
\frac{2}{1-\alpha^2}\,\quad (b>0),
\\[2pt]
\frac{2\alpha^2}{\alpha^2-1}\,\quad (b<0).
\end{cases}
\end{equation}
The relation between entropy and temperature in the near-extremal limit is \cite{Huijse:2011ef}
\begin{equation}
S\propto T^\frac{2-\theta}{\mathsf{z}}.
\label{eq:ST}
\end{equation}

A constraint for the hyperscaling-violating geometry is the null energy condition (NEC), which gives \cite{Dong:2012se}: $(\mathsf{d}-\theta)(\mathsf{d}(\mathsf{z}-1)-\theta)\geq 0$ and $(\mathsf{z}-1)(\mathsf{d}+\mathsf{z}-\theta)\geq 0$, where $\mathsf{d}$ is the spatial dimension. For $\mathsf{d}=2$ and $\mathsf{z}=1$, the NEC gives $(2-\theta)\theta\leq 0$, which is always satisfied for our solution. The Gubser criterion gives a stronger constraint on the parameter $\alpha$ (or $\theta$).

\begin{figure}
\centering
  \includegraphics[]{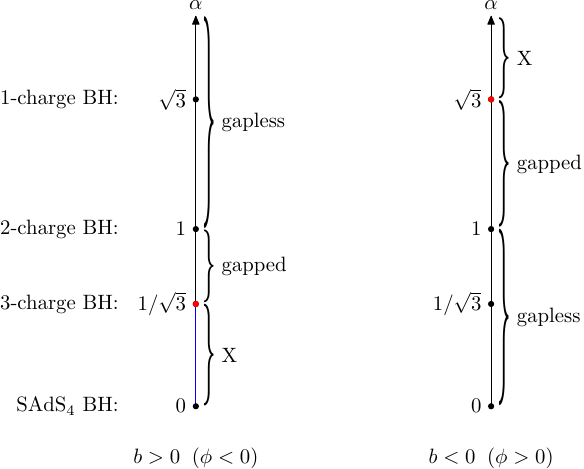}
  \caption{\label{fig:IR-AdS4} Classification of the one-parameter family of neutral solutions in AdS$_4$ according to their IR geometries. Special cases belonging to supergravity are marked. SAdS$_4$ denotes the planar Schwarzschild-AdS$_4$ black hole. ``X'' denotes that in this range of parameter space, the hyperscaling-violating geometry violates the Gubser criterion. In the blue interval ($0\leq\alpha\leq 1/\sqrt{3}$), a class of charged black hole solutions exists, and the IR geometries are changed by the gauge field.}
\end{figure}

For the Einstein-scalar system, we analyze the IR geometries according to \cite{Charmousis:2010zz}; see also \cite{Kiritsis:2015oxa,Ren:2019lgw}. We need to identify the leading exponential in the scalar potential~\eqref{eq:3-exp-V}. Classification by IR geometries is summarized in figure~\ref{fig:IR-AdS4}. If $b>0$ ($\phi<0$), the leading term in $V(\phi)$ in the IR is the first term. When $\alpha>1$, the extremal geometry is at $T\to 0$, and the spectrum is gapless. When $1/\sqrt{3}<\alpha\leq 1$, the extremal geometry is at $T\to\infty$, and the spectrum is gapped. When $0<\alpha\leq 1/\sqrt{3}$, it violates the Gubser criterion, and thus is unacceptable holographically.
If $b<0$ ($\phi>0$), the leading term in $V(\phi)$ in the IR is the last term. The potential is invariant under the transformation $\alpha\to 1/\alpha$ and $\phi\to -\phi$, which relates the $b>0$ and $b<0$ classifications. Figure~\ref{fig:IR-AdS4} clearly shows the special positions of STU supergravity in the one-parameter family of the scalar potential.

When the Gubser criterion is satisfied, the solution~\eqref{eq:solution} describes the ground state of holographic superconductors at zero density. To support this claim, we need to find finite temperature solutions under the following conditions: (i) The boundary condition for the scalar is sourceless and compatible with \eqref{eq:solution}. (ii) The hairy solution has lower free energy below a critical temperature. (iii) When we take the extremal limit, the finite temperature solution approaches the ground state solution. The FG expansion of $\phi$ in the solution~\eqref{eq:solution} is
\begin{equation}
\phi=-\frac{2\alpha b}{(1+\alpha^2)L^2}\tilde{z}-\frac{\alpha(1-\alpha^2)b^2}{(1+\alpha^2)^2L^4}\tilde{z}^2+\cdots,
\end{equation}
where $\tilde{z}$ is the FG radial coordinate. Therefore, the solution is compatible with a boundary condition $\phi_b=\kappa\phi_a$ due to a double-trace deformation with parameter
\begin{equation}
\kappa=\frac{1-\alpha^2}{2(1+\alpha^2)L^2}b\,.
\end{equation}
In the gapless cases, we have $\kappa<0$. In the next section, we will consider the gapless cases and one special case ($\alpha=1$) of gapped cases. The gapped cases need further investigation.

A way to analyze the stability of the ground state was studied in \cite{Faulkner:2010fh}.
The potential of the scalar field can be written in terms of a superpotential \cite{Lu:2014ida}:
\begin{equation}
V(\phi)=P'(\phi)^2-\frac{3}{4}P^2,\qquad P(\phi)=\frac{2\sqrt{2}}{(1+\alpha^2)L}\Bigl(e^{\frac{\alpha}{2}\phi}+\alpha^2 e^{-\frac{1}{2\alpha}\phi}\Bigr).
\end{equation}
The superpotential for small $\phi$ is
\begin{equation}
P(\phi)=\frac{2\sqrt{2}}{L}\biggl(1+\frac{1}{8}\phi^2-\frac{1-\alpha^2}{48\alpha}\phi^3+\mathcal{O}(\phi^4)\biggr).\label{eq:superP}
\end{equation}
According to \cite{Faulkner:2010fh}, with \eqref{eq:superP}, the stability of the system can be analyzed by an off-shell potential
\begin{equation}
\mathcal{V}(\phi_a)=\frac{1}{2}\biggl(\kappa\phi_a^2+\frac{2s_c}{3}|\phi_a|^3\biggr),\qquad s_c=-\frac{1-\alpha^2}{4\alpha}\text{sgn}(b).\label{eq:calV}
\end{equation}
We need $s_c>0$ for $\mathcal{V}$ to have a global minimum, and this corresponds to the gapless cases of our system.

\section{Superconducting phase}
\label{sec:SC}
The normal phase is simply the planar Schwarzschild-AdS$_4$ black hole. The AC conductivity calculated from this background geometry is a constant \cite{Herzog:2007ij}. In the following, we focus on the superconducting phase.

We numerically solve the Einstein-scalar system~\eqref{eq:ES} with the potential~\eqref{eq:3-exp-V} and obtain hairy black holes at finite temperature. This is a boundary value problem and the numerical technique was described in detail in \cite{Hartnoll:2008kx}. The metric ansatz is
\begin{equation}
ds^2=\frac{1}{z^2}\left(-g(z)e^{-\chi(z)}dt^2+\frac{dz^2}{g(z)}+dx^2+dy^2\right).\label{eq:ansatz}
\end{equation}
The AdS boundary is at $z=0$, and the horizon is at $z=z_h$, which can be fixed to be $z_h=1$ by scaling symmetries. The superconducting phase corresponds to a neutral black hole with scalar hair, which exists below a critical temperature. First we solve the background geometry, and then we perturb the system to calculate the conductivity.

The equations of motion for the metric and the scalar field are
\begin{align}
& g'-\left(\frac{\chi'}{2}+\frac{3}{z}\right) g-\frac{1}{2z}V(\phi)=0\,,\label{eq:eom-g}\\
& \chi'-\frac{1}{2}z\phi'^2=0\,,\label{eq:eom-chi}\\
& \phi ''+\left(\frac{g'}{g}-\frac{\chi'}{2}-\frac{2}{z}\right) \phi' -\frac{1}{z^2 g}V'(\phi)=0\,.\label{eq:eom-phi}
\end{align}
Near the horizon $z=1$, the asymptotic behavior of the functions is
\begin{align}
g &=\bar{g}_1(1-z)+\bar{g}_2(1-z)^2+\cdots,\\
\chi &=\chi_h+\bar{\chi}_1(1-z)+\cdots,\\
\phi &=\phi_h+\bar{\phi}_1(1-z)+\cdots,
\end{align}
where $\chi_h$ and $\phi_h$ can be used to express other coefficients. The dimensionless temperature and entropy density are given by
\begin{equation}
\tilde{T}=\frac{T}{-\kappa}=\frac{\bar{g}_1e^{-\chi_h/2}}{4\pi(-\kappa)},\qquad \tilde{s}=\frac{s}{(-\kappa)^2}=\frac{4\pi}{\kappa^2}\,.
\end{equation}
Near the AdS boundary $z=0$, the asymptotic behavior of the functions is
\begin{align}
g &= 1+\frac{1}{4}\phi_a^2z^2+g_3z^3+\cdots,\label{eq:fbdy}\\
e^{-\chi}g &= 1-\frac{1}{2}m_0 z^3+\cdots,\label{eq:hbdy}\\
\phi &= \phi_a z+\phi_b z^2+\cdots,\label{eq:phibdy}
\end{align}
where $-m_0/2=g_3-(2/3)\phi_a\phi_b$. Moreover, $m_0=(2/3)Ts$ can be derived by the radially conserved quantity~\eqref{eq:noether} below. The horizon value $\chi_h$ is chosen such that $\chi=0$ at the boundary, which is implemented by rescaling the time. Thus, integrating out from the horizon to infinity gives a map
\begin{equation}
\phi_h\mapsto (\phi_a,\phi_b,g_3)\,.
\end{equation}
Upon imposing the boundary condition~\eqref{eq:kappa}, we obtain a one-parameter family of solutions.\footnote{The equations~\eqref{eq:eom-g}--\eqref{eq:eom-phi} have four integration constants. At the horizon, there are two constraints, $g=0$ and regularity of $\phi$. At the AdS boundary, there are two constraints, normalization of time and $\phi_b=\kappa\phi_a$, where $\kappa$ is a parameter.} We can think of this parameter as being the temperature of the theory at a fixed $\kappa$.

\begin{figure}
\centering
\includegraphics[height=0.305\textwidth]{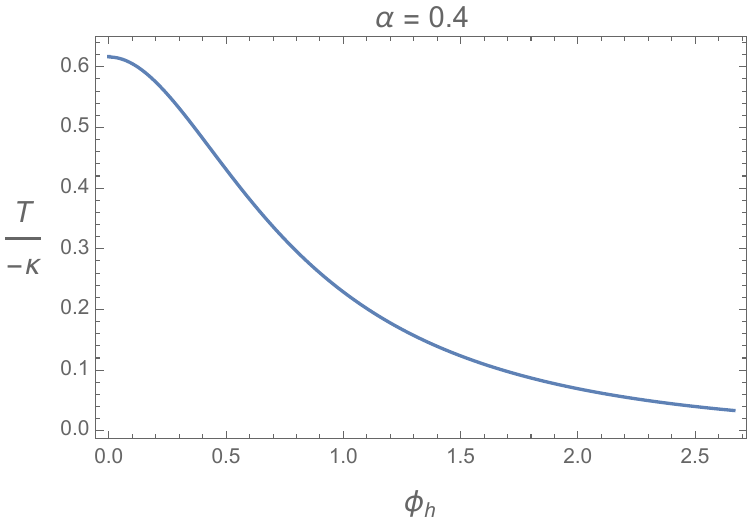}\qquad
\includegraphics[height=0.305\textwidth]{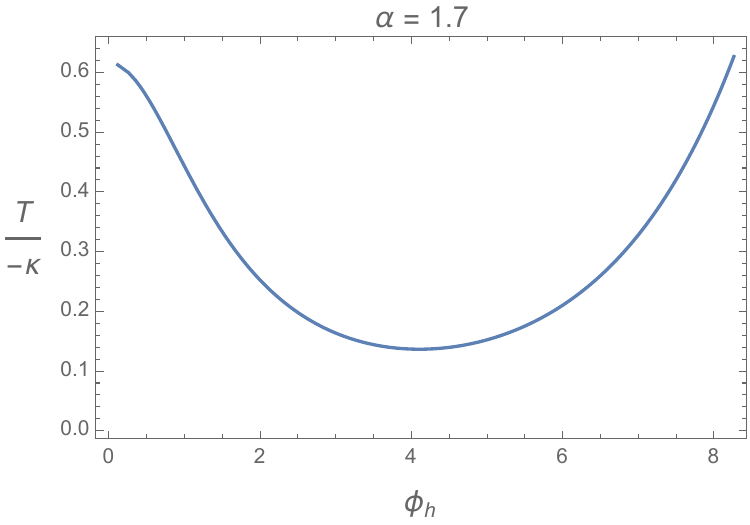}
\caption{\label{fig:Tphih} The temperature as a function of $\phi_h$. Left panel: when $\alpha=0.4$, the extremal limit is at $\tilde{T}\to 0$. Right panel: when $\alpha=1.7$, the extremal limit is at $\tilde{T}\to\infty$.}
\end{figure}

To have an instability, we take $\kappa<0$. The temperature changes as we change $\kappa$, and the extremal limit is at $\kappa\to -\infty$. The operator dual to the scalar field condenses without being sourced below a critical temperature $T_c$. In addition, we expect that the finite temperature geometry approaches the ground state geometry~\eqref{eq:solution} as we take the extremal limit. When the system is gapless, the extremal limit is at $T\to 0$; when the system is gapped, the extremal limit is at $T\to\infty$ \cite{Charmousis:2010zz}. In the latter case, see \cite{Aprile:2012sr} for an example of holographic superconductors with a retrograde condensate as an unstable branch. We take two values of $\alpha=0.4$ and $1.7$ and show the temperature as a function of $\phi_h$ in figure~\ref{fig:Tphih}. While the full parameter space has rich properties, in the following we take a specific value $\alpha=0.4$ to demonstrate that the extremal limit approaches the ground state solution. Here $\alpha=0.4$ is a typical value in the gapless case $\alpha\in (0,1/\sqrt{3}\approx 0.577)$.

\begin{figure}
\centering
\includegraphics[height=0.305\textwidth]{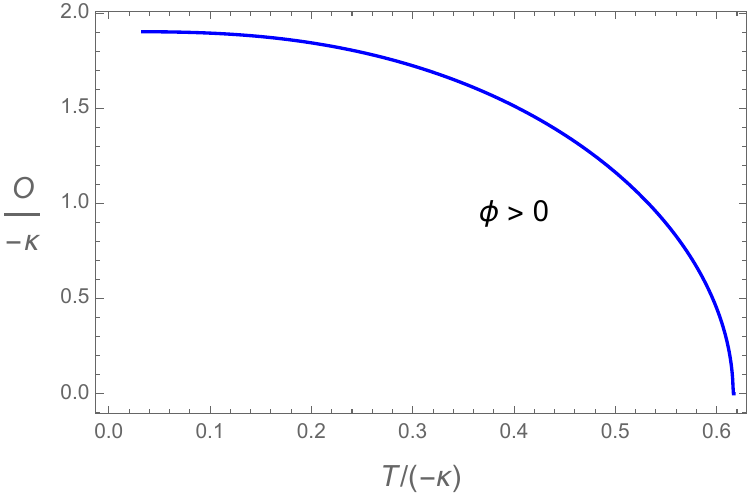}\qquad
\includegraphics[height=0.3\textwidth]{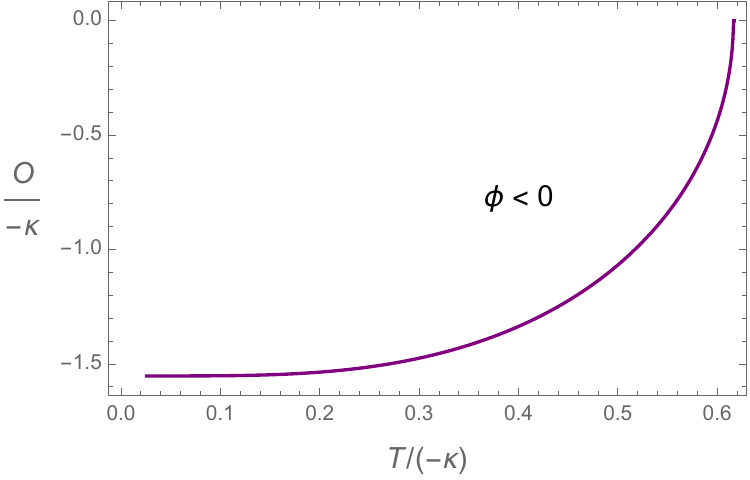}\\[5pt]
\includegraphics[height=0.3\textwidth]{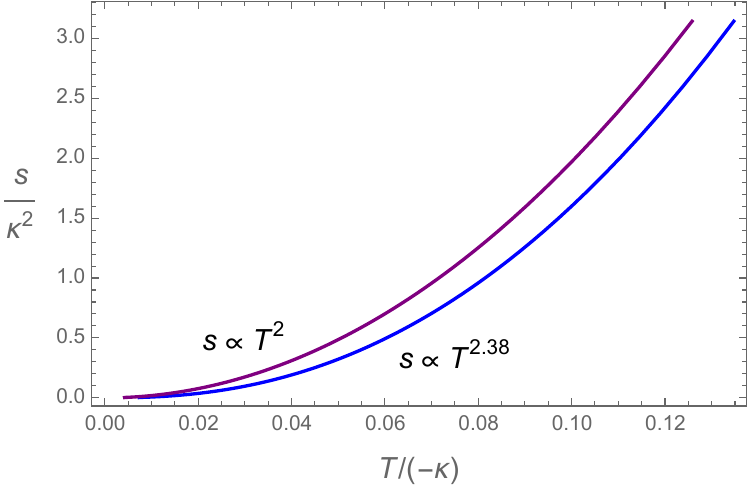}\;\,\quad
\includegraphics[height=0.3\textwidth]{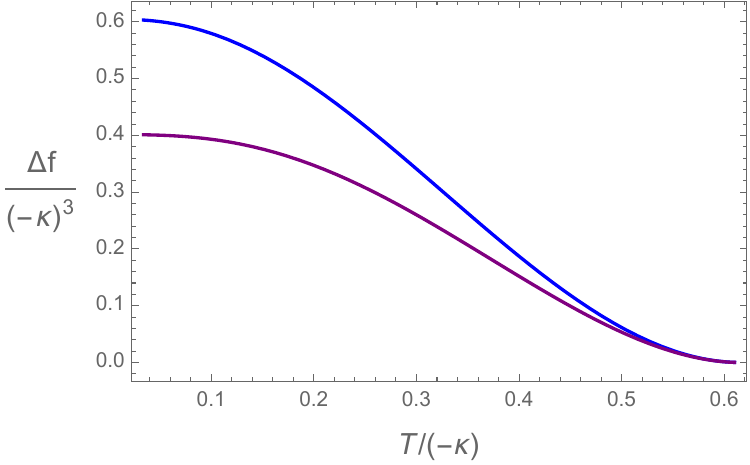}
\caption{\label{fig:OSFT} We choose $\alpha=0.4$. Upper left panel: the condensate of the operator $\langle{\cal O}\rangle$ as a function of temperature for the $\phi>0$ solution. Upper right panel: the condensate of the operator $\langle{\cal O}\rangle$ as a function of temperature for the $\phi<0$ solution. Lower left panel: entropy density as a function temperature at low temperatures for the two solutions. Lower right panel: difference of the free energy density between the black holes with and without scalar hair for the two solutions. The blue ($\phi>0$) curve has lower free energy, and gives the expected hyperscaling-violating geometry in the IR in the extremal limit.}
\end{figure}

We find that there are two solutions of hairy black holes. One has $\phi>0$ and the other has $\phi<0$. The condensates of $\langle{\cal O}\rangle=\phi_a$ are plotted in figure~\ref{fig:OSFT}. We will show that the extremal limit of the $\phi>0$ solution gives the expected hyperscaling-violating geometry in the IR. By fitting the curve as $\langle{\cal O}\rangle=a(1-T/T_c)^b$, where $T_c$ is the critical temperature, we obtain $b\approx 0.5$ as $T\to T_c$. This implies that a second-order phase transition occurs. We find
\begin{equation}
\langle{\cal O}\rangle\approx 4.5T_c(1-T/T_c)^{1/2},
\end{equation}
where $T_c/(-\kappa)\approx 0.616$. We use the entropy as a power law in temperature in the near-extremal limit to distinguish different solutions of hairy black holes. Recall that the Schwarzschild-AdS$_4$ black hole has $S\propto T^2$. If the entropy as a function of temperature has the behavior \eqref{eq:ST}, then the IR geometry is a hyperscaling-violating geometry. If we obtain $S\propto T^2$, then the IR geometry is AdS$_4$. In figure~\ref{fig:OSFT}, the lower left panel shows the entropy density as a function temperature at low temperatures for the two solutions.

The free energy is calculated by the renormalized on-shell action as $F/T=S_E+S_\text{ct}$, where $S_E$ is the Euclidean action and $S_\text{ct}$ is boundary counterterms. The free energy density is the free energy per volume: $\mathsf{f}=F/V_2$. The Euclidean on-shell action is a total derivative, and is evaluated as
\begin{equation}
S_E=\int d^3x\int_1^{0}dz\biggl(\frac{2}{z^3}ge^{-\chi/2}\biggr)'=\int d^3x\,\frac{2}{z^3}ge^{-\chi/2}\biggr|_{z=0}.
\end{equation}
The boundary terms are \cite{Faulkner:2010gj}\footnote{The boundary terms for the alternative quantization were given in \cite{Hartnoll:2008kx}. For the metric~\eqref{eq:ansatz}, the unit normal vector is $n=-z\sqrt{g}\,\partial_z$ pointing out of the AdS boundary, and the extrinsic curvature is $K_{\mu\nu}=\nabla_{(\mu}n_{v)}$. To compare the expressions in \cite{Faulkner:2010gj} and this work, we need to  rescale the scalar field by $\psi\to\phi/\sqrt{2}$.}
\begin{equation}
S_\text{ct} =\int d^3x\sqrt{\gamma}\biggl(-2K+4-\phi n^\mu\partial_\mu\phi-\frac{1}{2}\phi^2\biggr)\biggr|_{z=0}-\int d^3x\,(\phi_a\phi_b-W(\phi_a))\,,
\end{equation}
which includes the Gibbons-Hawking term, counterterms for the alternative quantization, and finite terms due to a general boundary condition $\phi_b=W'(\phi_a)$. The finite terms are chosen such that the following thermodynamic law is satisfied:
\begin{equation}
\delta\mathsf{f}=-s\delta T-(\phi_b-W'(\phi_a))\delta\phi_a\,.
\end{equation}
For the double-trace deformation, we have $W(\phi_a)=(1/2)\kappa\phi_a^2$. The free energy density is give by
\begin{equation}
\mathsf{f}=-\frac{1}{2}m_0+\frac{1}{6}\phi_a\phi_b=g_3-\frac{1}{2}\phi_a\phi_b\,.
\end{equation}
In figure~\ref{fig:OSFT}, the lower right panel shows the difference of the free energy density between the black holes with and without scalar hair at the same temperature for the two solutions: $\Delta\mathsf{f}=\mathsf{f}_0-\mathsf{f}$, where $\mathsf{f}_0$ is the free energy density for the Schwarzschild-AdS$_4$ black hole, and $\mathsf{f}$ is the free energy density for a hair black hole. The $\phi>0$ solution has lower free energy, and gives the expected hyperscaling-violating geometry in the IR in the extremal limit. For the ground state, we have $m_0=0$, and the free energy density is consistent with the off-shell potential~\eqref{eq:calV} $\mathcal{V}=(1/6)\phi_a\phi_b$.

To calculate the AC conductivity, we perturb the system by an alternating electric field
along the $x$ direction, which is achieved by adding a vector potential $\delta A_x=e^{-i\omega t}A_x(z)$. For zero density systems, the gauge field fluctuations decouple from metric fluctuations, and thus are significantly simplified. The perturbation equation for $A_x$ is given by
\begin{equation}
A_x''+\biggl(\frac{g'}{g}-\frac{\chi'}{2}+\frac{Z'(\phi) \phi'}{Z(\phi)}\biggr) A_x'+\biggl(\frac{\omega^2}{g^2}e^\chi-\frac{\mathcal{W}(\phi)}{z^2 g Z(\phi)}\biggr) A_x=0\,.
\label{eq:Ax}
\end{equation}
We choose the functions $Z$ and $\mathcal{W}$ as
\begin{equation}
Z(\phi)=1,\qquad \mathcal{W}(\phi)=\frac{1}{L^2}\sinh^2\Bigl(\frac{1+\alpha^2}{4\alpha}\phi\Bigr),
\label{eq:ZW}
\end{equation}
motivated by the special case $\alpha=1$ being a consistent truncation of M-theory. The asymptotic behavior of $A_x$ near the horizon is
\begin{equation}
A_x =(1-z)^{-\frac{i\omega}{4\pi T}}[\bar{a}_0+\bar{a}_1(1-z)+\cdots]\,,
\end{equation}
where we have chosen the infalling boundary condition to calculate the retarded Green's function. The asymptotic behavior near the AdS boundary is
\begin{equation}
A_x = \mathcal{A}+\mathcal{B}z+\cdots.\label{eq:sig0}
\end{equation}
The current-current correlation function and the conductivity are obtained by
\begin{equation}
G=\frac{\cal B}{\cal A},\qquad \sigma=\frac{G}{i\omega}\,.
\end{equation}

\begin{figure}
\centering
\centering
\includegraphics[height=0.295\textwidth]{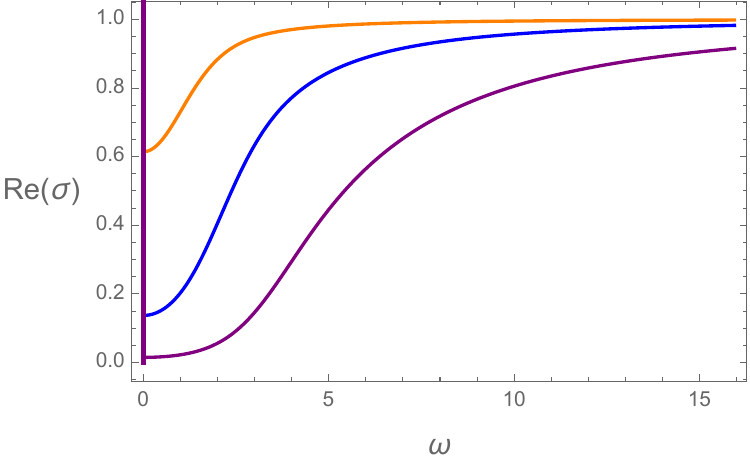}\quad
\includegraphics[height=0.295\textwidth]{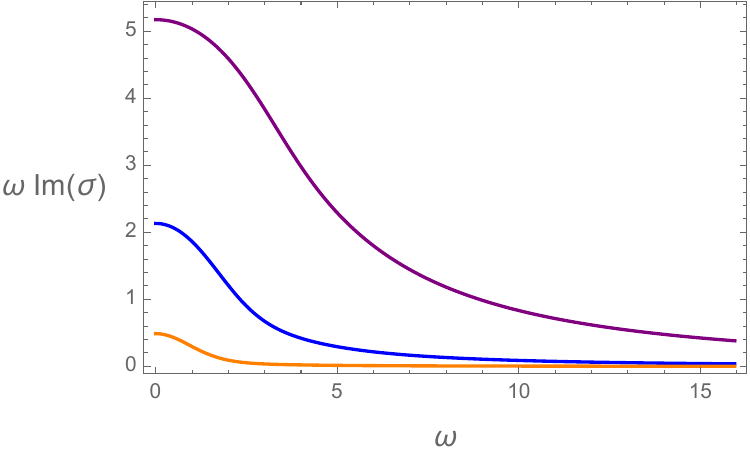}
\caption{\label{fig:sig} The behavior of conductivity as a function of frequency. The right panel shows that $\omega\text{Im}(\sigma)$ approaches to a constant as $\omega\to
0$, which implies that there are a pole in Im($\sigma$) and a delta function in Re($\sigma$) at $\omega=0$. The orange, blue, and purple curves correspond to $T/T_c=0.5$, $0.2$, $0.1$ ($\kappa=-0.8$, $-2.2$, $-5$), respectively.}
\end{figure}

The behavior of $\omega{\rm Im}(\sigma)$ is plotted in figure~\ref{fig:sig}, from which we can see that there is a pole in the imaginary part of the conductivity in the superconducting phase. The Kramers-Kronig relation implies a relation between the real and imaginary parts. In particular, we have
\begin{equation}
\frac{i}{\omega+i\epsilon}=\mathcal{P}\frac{i}{\omega}+\pi\delta(\omega)\,.
\end{equation}
Therefore, there is a delta function at $\omega=0$ in the real part, which is the feature of the superconducting phase at $T<T_c$. Recall that the conductivity is a constant in the normal phase. The Ferrell-Glover-Tinkham sum rule states that $\int{\rm Re}(\sigma)d\omega$ is independent of the temperature, which implies that the missing area is exactly compensated by the strength of the superconducting delta function:
\begin{equation}
\lim_{\omega\to 0}\omega{\rm
Im}[\sigma(\omega)]=\frac{2}{\pi}\int_{0^+}^\infty(1-\text{Re}[\sigma(\omega)])d\omega\,.
\end{equation}
This has been proved in holography in \cite{Gulotta:2010cu}. We have used the sum rule as a cross-check for our numerical calculations.

\section{A holographic superconductor from M-theory}
\label{sec:MT}
Interestingly, the $\alpha=1$ case of our system \eqref{eq:action} with \eqref{eq:3-exp-V} and \eqref{eq:ZW} can be embedded in M-theory. We find an analytic solution of the AC conductivity for this holographic superconductor at zero density. The model is \cite{Donos:2011ut,Aprile:2012sr}
\begin{equation}
\mathcal{L}=R-{1\over 2}(\partial\phi)^2+\frac{2}{L^2}(\cosh\phi+2)-\frac{1}{4}F^2-\frac{1}{L^2}\sinh^2\Bigl(\frac{\phi}{2}\Bigr)A^2\,.
\label{eq:action2}
\end{equation}
When $A_\mu=0$, the background geometry is the same as the 2-charge black hole in AdS$_4$, which is the $\alpha=1$ case of \eqref{eq:solution}. The metric is
\begin{equation}
ds^2=\frac{r(r-b)}{L^2}(-dt^2+d\vec{x}^2)+\frac{dr^2}{r(r-b)}.
\label{eq:2-charge-bh}
\end{equation}
Since both $g_{tt}$ and $g^{rr}$ have a single root, the geometry apparently has a finite temperature
\begin{equation}
T=\frac{|b|}{4\pi L^2}.
\end{equation}
The dimensionless temperature $\tilde{T}\equiv T/(-\kappa)$ is infinity, implying that the near-extremal solution is in an unstable branch. The Kretschmann scalar is
\begin{equation}
R_{\mu\nu\rho\sigma}R^{\mu\nu\rho\sigma}=\frac{3 \left(32 r^4-64 b r^3+40 b^2 r^2-8 b^3 r+b^4\right)}{4 L^4 (r-b)^2 r^2},
\end{equation}
which indicates that there is a spacetime singularity at $r=b$ or $r=0$. This is consistent with the statement that the background geometry~\eqref{eq:2-charge-bh} describes the ground state of a holographic superconductor. As pointed out in \cite{Aprile:2012sr}, the ground state of this holographic superconductor is in a confined cohesive phase.

To calculate the conductivity, we perturb the system by $e^{-i\omega t}A_x(r)$ around the background geometry~\eqref{eq:2-charge-bh}. The perturbation equation for $A_x(r)$ is
\begin{equation}
A_x''+\frac{2r-b}{r(r-b)}A_x'+\frac{4\omega^2L^4-b^2}{4r^2(r-b)^2}A_x=0\,.
\end{equation}
The solution of $A_x(r)$ is
\begin{equation}
A_x=C_1\biggl(\frac{r-b}{r}\biggr)^\frac{\sqrt{b^2-4\omega^2L^4}}{2b}+C_2\biggl(\frac{r-b}{r}\biggr)^{-\frac{\sqrt{b^2-4\omega^2L^4}}{2b}}.
\end{equation}
When $2\omega L^2>|b|$, the first term with the $\omega\to\omega+i\epsilon$ prescription describes infalling wave in the IR for either $b>0$ or $b<0$. Thus the infalling boundary condition requires $C_2=0$. The Green's function is given by
\begin{equation}
  G(\omega)=-\frac{1}{2L^2}\Bigl(\sqrt{b^2-4\omega^2L^4}\Bigr).
\label{eq:G-alpha-1}
\end{equation}
There are branch cuts $4\omega^2L^4>b^2$ on the real axis of the complex $\omega$ plane. We expected that the branch cuts become dense poles in the lower-half plane as the system is away from the extremal limit.

The conductivity is obtained by $\sigma=G/i\omega|_{\omega\to\omega+i\epsilon}$, and the real part is
\begin{equation}
\text{Re}[\sigma(\omega)]=\frac{\pi |b|}{2L^2}\delta(\omega)+\theta(4\omega^2L^4-b^2)\frac{\sqrt{4\omega^2L^4-b^2}}{2\omega L^2},
\label{eq:sigma4-2}
\end{equation}
where $\theta(x)$ is the Heaviside step function. There are a delta function at $\omega=0$, and a gap in $0<2\omega L^2<b$. This analytic solution is among the very rare analytic solutions of two-point functions calculated from the AdS/CFT for nonconstant curvature geometries.

For zero density systems that are not superconducting, the conductivity can also have a delta function at $\omega=0$ at zero temperature (not at $T>0$). As an example, a conductivity with a delta function and a hard gap was solved in \cite{DeWolfe:2012uv} for the Coulomb branch solution in AdS$_5$, which is the 1-charge black hole in AdS$_5$.
It was pointed out that this type of delta function at $\omega=0$ together with the hard gap is closely related to superconductivity \cite{DeWolfe:2012uv}. However, to have a superconducting delta function, we need a spontaneous breaking of the U(1) symmetry. Zero density systems can share the same background geometry while having different gauge field fluctuations. For the system~\eqref{eq:action2} with a $A^2$ term, the delta function in \eqref{eq:sigma4-2} is due to superconductivity.

\section{A more general solution}
\label{sec:general}
A more general solution can be obtained by taking the nontrivial neutral limit of the following Einstein-Maxwell-dilaton-axion (EMDA) system with a six-exponential potential. Then this neutral system can be promoted to a holographic superconductor at zero density.
The action with axion (massless scalar) fields $\chi_i$ is
\begin{equation}
\mathcal{L}=R-\frac14 e^{-\alpha\phi}F^2-\frac12(\partial\phi)^2-V(\phi)
-\frac{1}{2}\sum_{i=1}^{2}(\partial\chi_i)^2,\label{eq:axions4}
\end{equation}
where $\chi_i=a x_i$ satisfies the equation of motion of $\chi_i$ ($i=1,2$). This system was used as a simple way to introduce momentum dissipation, since $a x_i$ breaks the translation symmetry \cite{Andrade:2013gsa,Gouteraux:2014hca}. For more general axion models, see \cite{Baggioli:2021xuv} for a review. Let $V_\alpha(\phi)$ be the three-exponential potential~\eqref{eq:3-exp-V}, and we consider the following six-exponential potential \cite{Ren:2019lgw}\footnote{Other ways to derive a six-exponential potential can be found in \cite{Anabalon:2012ta,Feng:2013tza}, and the corresponding black hole thermodynamics is studied in \cite{Anabalon:2019tcy}.}
\begin{equation}
V(\phi)=(1+\beta)V_\alpha(\phi)-\beta V_{-\alpha}(\phi)\,.
\label{eq:6-exp-V}
\end{equation}
The solution of the metric $g_{\mu\nu}$, gauge field $A_\mu$, and dilaton field $\phi$ is
\begin{align}
& ds^2 =-f(r)\tsp dt^2+\frac{1}{f(r)}\tsp dr^2+U(r)\tsp d\vec{x}^2\,,\label{eq:sol4}\\
& A =2\sqrt{\frac{bc}{1+\alpha^2}}\left(\frac{1}{r_h}-\frac{1}{r}\right)dt\,,\qquad
e^{\alpha\phi}=\left(1-\frac{b}{r}\right)^\frac{2\alpha^2}{1+\alpha^2},
\end{align}
with
\begin{equation}
\begin{split}
f &=\left(-\frac{a^2}{2}-\frac{c}{r}\right)\left(1-\frac{b}{r}\right)^\frac{1-\alpha^2}{1+\alpha^2}+\frac{1+\beta}{L^2}r^2\biggl(1-\frac{b}{r}\biggr)^\frac{2\alpha^2}{1+\alpha^2}\\
&\hspace{0.1\textwidth} -\frac{\beta}{L^2}\,r^2\biggl(1+\frac{1-3\alpha^2}{1+\alpha^2}\,\frac{b}{r}+\frac{(1-\alpha^2)(1-3\alpha^2)}{(1+\alpha^2)^2}\,\frac{b^2}{r^2}\biggr)\biggl(1-\frac{b}{r}\biggr)^{\frac{1-\alpha ^2}{1+\alpha ^2}},\\
U &=r^2\biggl(1-\frac{b}{r}\biggr)^\frac{2\alpha^2}{1+\alpha^2}.\label{eq:fsol4}
\end{split}
\end{equation}
The solution has parameters $a$, $b$ and $c$ in addition to $\alpha$ and $\beta$. The curvature singularity is at $r=0$ and $r=b$. In the trivial neutral limit $b=0$, the scalar field $\phi$ vanishes. In the nontrivial neutral limit $c=0$, the scalar field $\phi$ is nonzero and we can obtain a finite temperature black hole for either $a\neq 0$ or $\beta\neq 0$.

As a remark, we can derive the potential~\eqref{eq:6-exp-V} in the following way. Suppose we have a solution without momentum dissipation, can we obtain a solution with momentum dissipation by simply adding additional terms in the blackening factor $f(r)$? It is impossible in general. However, if we demand that the answer is yes, it imposes a strong constraint to the system. As a consequence, the most general potential under this condition is \eqref{eq:6-exp-V} \cite{Ren:2019lgw}. A special case is the three-exponential potential \eqref{eq:3-exp-V} when $\beta=0$.

We gain more insights on STU supergravity solutions by comparing the charged and neutral black holes with scalar hair. The charged black hole solutions \cite{Gao:2004tu,Gouteraux:2014hca} are given by \eqref{eq:axions4}--\eqref{eq:fsol4} with $a=\beta=0$. They have a nontrivial neutral limit by taking $c=0$, which gives \eqref{eq:solution}. These solutions are characterized by their IR geometries in the extremal limit. The charged solution requires $b>0$, which is not necessary for the neutral solution. To compare them, we assume $b>0$ in the following.
\begin{itemize}
\item The charged solution has an extremal limit only for $0\leq\alpha\leq 1/\sqrt{3}$. When $0\leq\alpha< 1/\sqrt{3}$, the IR geometry is AdS$_2\times\mathbb{R}^2$. When $\alpha=1/\sqrt{3}$, the IR geometry is conformal to AdS$_2\times\mathbb{R}^2$. When $\alpha>1/\sqrt{3}$, the extremal limit becomes neutral \cite{Ren:2019lgw}.\footnote{As special cases, this peculiar behavior was noticed by \cite{DeWolfe:2012uv} for the 1-charge black hole in AdS$_5$, and \cite{Kiritsis:2015oxa} for the 1-charge and 2-charge black holes in AdS$_4$. This property is for the particular class of analytic solutions, not for the system in general. We can numerically add one more parameter to the system.}

\item The neutral solution is already at the extremal limit for $\alpha>1/\sqrt{3}$. When $0<\alpha\leq 1/\sqrt{3}$, the neutral solution violates the Gubser criterion and thus is unacceptable holographically. Adding other matter fields can change the IR geometry.
\end{itemize}
The joining point $\alpha=1/\sqrt{3}$ is the 3-charge black hole in AdS$_4$, which is also called the Gubser-Rocha model \cite{Gubser:2009qt}.\footnote{The Gubser criterion and the Gubser-Rocha model were completely unrelated.} The two endpoints $\alpha=0$ and $\alpha\to\infty$ are black holes without scalar hair.

For the finite density system, we can also numerically construct a more general black hole solution by tuning the double-trace deformation parameter $\kappa$. Namely, the numerical solution for black holes constructed in section~\ref{sec:SC} is the zero density section of a higher dimensional parameter space. If the gauge field is irrelevant in the IR, a finite density system shares the same IR geometry as a zero density system.

\section{Discussion}
\label{sec:sum}
We have constructed holographic superconductors at zero density by promoting a neutral Einstein-scalar system to holographic superconductors, as an analytic realization of \cite{Faulkner:2010gj}. The ground state has an analytic solution, and its IR is a hyperscaling-violating geometry. We take advantage of the fact that the following systems share the same background geometry at zero density when they intersect:
\begin{itemize}
\item[(a)] a nontrivial neutral limit of an EMD system with a three-exponential potential,

\item[(b)] special cases of STU supergravity,

\item[(c)] holographic superconductors at zero density.
\end{itemize}
As a specific example, we obtain an analytic solution of the AC conductivity for a holographic superconductor from M-theory. The finite temperature solutions are numerically constructed for a typical value of $\alpha$. By a six-exponential potential, we can analytically write down a finite temperature solution, and the parameter space is large. The full parameter space needs to be explored.

With the solution of the ground state, we can analytically solve the AC conductivity for general $\alpha\neq 1$, in addition to the $\alpha=1$ result \eqref{eq:sigma4-2}. We choose the functions $Z$ and $\mathcal{W}$ as \eqref{eq:ZW}, and perturb the system by $e^{-i\omega t}A_x(r)$. The general solution of $A_x(r)$ is
\begin{equation}
A_x(r)=C_1\tsp ze^{-z/2}{_1F_1}\biggl(1-\frac{i\gamma b}{8\omega L^2},\, 2,\,z\biggr)+C_2\tsp ze^{-z/2}U\left(1-\frac{i\gamma b}{8\omega L^2},\, 2,\,z\right),\label{eq:sol-Ax-alpha}
\end{equation}
where
\begin{equation}
\gamma\equiv\frac{1+\alpha^2}{1-\alpha^2},\qquad z\equiv\frac{2i\gamma\omega L^2}{b}\biggl(1-\frac{b}{r}\biggr)^{1/\gamma},
\end{equation}
and the functions ${_1F_1}$ and $U$ are solutions to the confluent hypergeometric equation. They have the following asymptotic behaviors\footnote{It is helpful to understand the functions ${_1F_1}$ and $U$ by recalling that the modified Bessel functions $I$ and $K$ are special cases of them:
\begin{equation*}
I_\nu(x)=\frac{2^{-\nu}}{\Gamma(\nu+1)}x^\nu e^{-x}{_1F_1}(1/2+\nu,1+2\nu,2x),\qquad K_\nu(x)=\sqrt{\pi}\,(2x)^\nu e^{-x}U(1/2+\nu,1+2\nu,2x).
\end{equation*}}
\begin{equation}
{_1F_1}(a,b,z)=1+\frac{a}{b}z+\cdots,\qquad z\to 0\,.
\end{equation}
\begin{equation}
U(a,b,z)=z^{-a}+\cdots,\qquad z\to\infty\,.
\end{equation}

To impose boundary conditions in the IR, we need to analyze the asymptotic behaviors of $A_x$. There are two distinctive cases:
\begin{itemize}
\item[(A)] $b>0$ and $\alpha<1$; or $b<0$ and $\alpha>1$. The IR is at $z\to 0$, where the boundary condition is normalizability. The solution of $A_x$ is \eqref{eq:sol-Ax-alpha} with $C_2=0$. The Green's function is
\begin{equation}
G(\omega)=i\omega-\frac{b}{\gamma L^2}-\biggl(i\omega+\frac{\gamma b}{8L^2}\biggr)
\frac{{_1F_1}\Bigl(2-\frac{i\gamma b}{8\omega L^2},\, 3,\, \frac{2i\gamma\omega L^2}{b}\Bigr)}{{_1F_1}\Bigl(1-\frac{i\gamma b}{8\omega L^2},\, 2,\, \frac{2i\gamma\omega L^2}{b}\Bigr)}.
\end{equation}

\item[(B)] $b>0$ and $\alpha>1$; or $b<0$ and $\alpha<1$. The IR is at $z\to -i\infty$, where the boundary condition is infalling wave. The solution of $A_x$ is \eqref{eq:sol-Ax-alpha} with $C_1=0$. The Green's function is
\begin{equation}
G(\omega)=i\omega-\frac{b}{\gamma L^2}-\biggl(i\omega+\frac{\gamma b}{8L^2}\biggr)
\frac{U\Bigl(2-\frac{i\gamma b}{8\omega L^2},\, 3,\, \frac{2i\gamma\omega L^2}{b}\Bigr)}{U\Bigl(1-\frac{i\gamma b}{8\omega L^2},\, 2,\, \frac{2i\gamma\omega L^2}{b}\Bigr)}.
\end{equation}
\end{itemize}
The conductivity is obtained by $\sigma=G/i\omega|_{\omega\to\omega+i\epsilon}$. In case~(A), the real part of the conductivity consists of an infinite number of delta functions, and thus, is gapped. In case~(B), the real part of the conductivity is gapless. In any case, the real part of the conductivity has a delta function at $\omega=0$, indicating superconductivity.

\acknowledgments
J.R. thanks Christopher Herzog, Elias Kiritsis, Li Li, Hong L\"{u}, and W\'{e}i W\'{u} for helpful discussions. This work was supported in part by the NSF of China under Grant No. 11905298.

\appendix
\section{Equations of motion and holographic renormalization}
\label{sec:HR}
Equations of motion from the action~\eqref{eq:action} with $p=0$ in $D$ dimensions are
\begin{align}
& R_{\mu\nu}=\frac{1}{2}\bigl(\tilde{T}_{\mu\nu}^A+\tilde{T}_{\mu\nu}^\phi\bigr),\\
& \nabla^{\mu}\left(Z(\phi)F_{\mu\nu}\right)-\mathcal{W}(\phi)A_\nu=0,\\
& \nabla^2\phi-V'(\phi)-\frac{Z'(\phi)}{4}F^2-\frac{\mathcal{W}'(\phi)}{2}A^2=0,
\end{align}
where
\begin{align}
\tilde{T}_{\mu\nu}^A &= Z(\phi)\biggl(F_{\mu\rho}F_{\nu}^{\;\,\rho}-\frac{1}{2(D-2)}g_{\mu\nu}F^2\biggr)+\mathcal{W}(\phi)A_\mu A_\nu,\\
\tilde{T}_{\mu\nu}^\phi &= \partial_\mu\phi\partial_\nu\phi+\frac{2}{D-2}g_{\mu\nu}V(\phi).
\end{align}

We can derive the free energy in a different approach, following \cite{Caldarelli:2016nni} to calculate the mass of the black holes in Einstein-scalar systems by holographic renormalization. In the Fefferman-Graham (FG) gauge, the AdS$_4$ metric is written in the form
\begin{equation}
ds^2=\frac{L^2}{\tilde{z}^2}(d\tilde{z}^2+g_{ij}(x,\tilde{z})dx^i dx^j)\,,
\label{eq:FG}
\end{equation}
where $\tilde{z}$ is the FG radial coordinate, and $g_{ij}$ is a three-dimensional metric, which raises/lowers the $i, j$ indices. The asymptotic behavior of the metric and scalar field near the AdS boundary is
\begin{align}
g_{ij}(x,\tilde{z}) &=g_{(0)ij}+\tilde{z}g_{(1)ij}+\tilde{z}^2g_{(2)ij}+\cdots,\\
\phi &=\phi_a \tilde{z}^{\Delta_-}(1+\cdots)+\phi_b \tilde{z}^{\Delta_+}(1+\cdots).\label{eq:FGbry}
\end{align}
The boundary condition for a multi-trace deformation is specified starting from the alternative quantization.  The single-trace source is written as 
\begin{equation}
J_\mathcal{F}=-L^2\phi_b-\mathcal{F}'(\phi_a)\,,
\label{eq:calF}
\end{equation}
where $\mathcal{F}(\phi_a)$ is a polynomial, and $J_\mathcal{F}=0$ specifies a sourceless condition. With a general boundary condition for the scalar field, we need to add an additional finite boundary term $S_\mathcal{F}$ to the renormalized on-shell action \cite{Caldarelli:2016nni}:
\begin{equation}
S'_\text{ren}=\lim_{\epsilon\to 0}(S_\text{bulk}+S_\text{GH}+S_\text{ct}+S_\mathcal{F})\,,
\end{equation}
where $S_\text{bulk}$ is the bulk action with the radial coordinate being integrated from the horizon to the cutoff $\tilde{z}=\epsilon$, and
\begin{align}
S_\text{GH} &=\int_{\tilde{z}=\epsilon} d^3x\sqrt{-\gamma}\,2K\,,\\
S_\text{ct} &=-\int_{\tilde{z}=\epsilon} d^3\sqrt{-\gamma}\Bigl(\frac{4}{L}+LR[\gamma]+\frac{1}{2L}\phi^2\Bigr),\\
S_\mathcal{F} &=\int_{\tilde{z}=\epsilon} d^3x\sqrt{-g_{(0)}}\Bigl(J_\mathcal{F}\phi_a+\mathcal{F}(\phi_a)\Bigr).
\end{align}
Its variation is
\begin{equation}
\delta S'_\text{ren}=\int d^3x\sqrt{-g_{(0)}}\Bigl(\frac{1}{2}\langle\mathcal{T}^{ij}\rangle\delta g_{(0)}^{ij}+\langle O\rangle\delta J_\mathcal{F}\Bigr).
\end{equation}
Consequently, the boundary stress tensor is given by \cite{Caldarelli:2016nni}
\begin{equation}
\langle\mathcal{T}^{ij}\rangle=3L^2g_{(3)}^{ij}+\Bigl(\mathcal{F}(\phi_a)-\phi_a\mathcal{F}'(\phi_a)\Bigr)g_{(0)}^{ij}\,.
\end{equation}
The energy density is given by $\epsilon=L^2\langle\mathcal{T}^{tt}\rangle$.
For a boundary condition corresponding to a double-trace deformation, the boundary stress tensor is given by
\begin{equation}
\langle\mathcal{T}^{ij}\rangle=3L^2g_{(3)}^{ij}+\frac{1}{2}\phi_a\phi_b g_{(0)}^{ij}\,.
\label{eq:Tij}
\end{equation}

For the system in section~\ref{sec:SC}, the free energy is calculated as follows. There is a radially conserved quantity \cite{Gubser:2009cg}
\begin{equation}
\mathcal{Q}=\frac{e^{\chi/2}}{z^2}\bigl(ge^{-\chi}\bigr)',\label{eq:noether}
\end{equation}
which satisfies $\mathcal{Q}'=0$. At the horizon, $\mathcal{Q}=-Ts$. At the AdS boundary, $\mathcal{Q}=3g_3-2\phi_a\phi_b$. This relates the horizon quantity $Ts$ to the boundary coefficients. For the metric~\eqref{eq:ansatz} with \eqref{eq:fbdy}--\eqref{eq:phibdy}, the relation between the coordinates $z$ and the FG radial coordinate $\tilde{z}$ is
\begin{equation}
z=\tilde{z}+\frac{1}{16}\phi_a^2\tilde{z}^3+\frac{1}{6}g_3\tilde{z}^4+\cdots.
\end{equation}
The metric in the FG expansion is
\begin{align}
& g_{ij}=\left(\begin{array}{ccc}
-1 & 0 & 0\\
0 & 1 & 0\\
0 & 0 & 1
\end{array}\right)
+\tilde{z}^2\left(\begin{array}{ccc}
\frac{1}{8}\phi_a^2 & 0 & 0\\
0 & -\frac{1}{8}\phi_a^2 & 0\\
0 & 0 & -\frac{1}{8}\phi_a^2
\end{array}\right)+\tilde{z}^3\left(\begin{array}{ccc}
-\frac{2}{3}(g_3-\phi_a\phi_b) & 0 & 0\\
0 & -\frac{1}{3}g_3 & 0\\
0 & 0 & -\frac{1}{3}g_3
\end{array}\right)+\cdots.\label{eq:phi-FG}
\end{align}
The solution~\eqref{eq:solution} is compatible with a boundary condition $\phi_b=\kappa\phi_a$ due to a double-trace deformation, and the choice of $\mathcal{F}(\phi_a)$ is
\begin{equation}
\mathcal{F}(\phi_a)=-\frac{1}{2}\kappa L^2\phi_a^2\,,\qquad \kappa=\frac{1-\alpha^2}{2(1+\alpha^2)L^2}b\,.
\end{equation}
The energy density calculated by \eqref{eq:Tij} and the free energy density are given by
\begin{align}
\epsilon &=-2g_3+\frac{3}{2}\phi_a\phi_b\,,\\
\mathsf{f} &=\epsilon-Ts=g_3-\frac{1}{2}\phi_a\phi_b\,.
\end{align}

As a remark, the solution~\eqref{eq:solution} is also compatible with a boundary condition $\phi_b=\tau\phi_a^2$ due to a triple-trace deformation:
\begin{equation}
\mathcal{F}(\phi_a)=-\frac{1}{3}\tau L^2\phi_a^3\,,\qquad \tau=-\frac{1-\alpha^2}{4\alpha}\,.
\end{equation}
The triple-trace deformation is marginal and the parameter $\tau$ is dimensionless.

\section{Neutral limit of the Gubser-Rocha model}
\label{sec:GR}

In the literature, the Gubser-Rocha model refers to as the 3-charge black hole in AdS$_4$ or the 2-charge black hole in AdS$_5$. In the Gubser-Rocha model, the specific heat and entropy are linear in the temperature near zero temperature \cite{Gubser:2009qt}. Another distinctive property is that the IR of the extremal geometry is conformal to AdS$_2\times\mathbb{R}^\mathsf{d}$ \cite{Gubser:2012yb}. The 3-charge black hole in AdS$_4$ is determined by
\begin{equation}
\mathcal{L}=R-\frac{1}{4}e^{-\frac{1}{\sqrt{3}}\phi}F^2-\frac{1}{2}(\partial\phi)^2+\frac{6}{L^2}\cosh\frac{\phi}{\sqrt{3}}\,.
\end{equation}
In the ordinary form, the solution of the metric $g_{\mu\nu}$, gauge field $A_\mu$, and dilaton field $\phi$ is
\begin{gather}
ds^2=e^{2\mathcal{A}}(-hdt^2+d\vec{x}^2)+\frac{e^{2\mathcal{B}}}{h}dr^2,\\
\mathcal{A}=\ln\frac{\bar{r}}{L}+\frac{3}{4}\ln\left(1+\frac{Q}{\bar{r}}\right),\qquad \mathcal{B}=-\mathcal{A},\qquad h=1-\frac{(\bar{r}_h+Q)^3}{(\bar{r}+Q)^3},\\
A=\frac{\sqrt{3Q(\bar{r}_h+Q)}}{L}\left(1-\frac{\bar{r}_h+Q}{\bar{r}+Q}\right)dt,\qquad
\phi=-\frac{\sqrt{3}}{2}\ln\left(1+\frac{Q}{\bar{r}}\right),
\end{gather}
where $\bar{r}_h$ is the horizon radius, $Q$ is a parameter related to the chemical potential, and $L$ is the AdS radius. The temperature of this black hole is
\begin{equation}
T=\left.\frac{|h'|e^{\mathcal{A}-\mathcal{B}}}{4\pi}\right|_{\bar{r}=\bar{r}_h}=\frac{3\sqrt{\bar{r}_h(\bar{r}_h+Q)}}{4\pi L^2}.
\end{equation}
The zero temperature black hole solution is at $\bar{r}_h=0$. The IR is at $\bar{r}=0$, which is a spacetime singularity. It will be cloaked by a horizon at finite temperature.

Apparently, the neutral limit of this black hole is at $Q=0$, which is nothing but the planar Schwarzschild-AdS black hole. However, a closer examination reveals that this is not the whole story. A different parametrization gives a neutral solution with a nontrivial dilaton profile.
Define
\begin{equation}
b=Q,\qquad c=(\bar{r}_h+Q)^3,\qquad r=\bar{r}+Q,
\end{equation}
and then set $c=0$, and then we obtain the solution~\eqref{eq:solution} in the $\alpha=1/\sqrt{3}$ case. As a neutral solution, $b$ is not necessarily positive. If $b>0$, the neutral solution (marginally) violates the Gubser criterion, and thus is unacceptable. If $b<0$, the neutral solution is the extremal limit of a finite temperature of black hole. These two cases are marked by red dots in figure~\ref{fig:IR-AdS4}.

\section{Kasner universe near the singularity}
\label{sec:KU}

At finite temperature, the metric describes a Kasner universe near the spacetime singularity when we dive inside the horizon of holographic superconductors \cite{Frenkel:2020ysx,Hartnoll:2020fhc}. We show that for the near-extremal geometry, we can obtain a simple Kasner universe.

We take the following IR metric of the AdS$_4$ near-extremal solution (we need to add appropriate coefficients before $dt^2$ and $dr^2$) \cite{Huijse:2011ef}:
\begin{equation}
ds^2=\frac{1}{r^2}\left(-fr^{-\frac{4(\mathsf{z}-1)}{2-\theta}}dt^2+r^\frac{2\theta}{2-\theta}\frac{dr^2}{f}+d\vec{x}^2\right),\qquad f=1-\Bigl(\frac{r}{r_h}\Bigr)^\frac{2(2+\mathsf{z}-\theta)}{2-\theta},
\end{equation}
where the horizon is at $r=r_h$, the spacetime singularity is at $r\to\infty$. The extremal solution can be obtained by sending $r_h\to\infty$. Since $r<r_h$ outside the horizon, the power in the blackening factor $f(r)$ must be positive as required by the Gubser criterion. The metric near the spacetime singularity can be written as
\begin{equation}
ds^2=-d\tau^2+c_t\tau^{2p_t}dt^2+c_x\tau^{2p_x}(dx^2+dy^2),\qquad \phi=-\sqrt{2}\,p_\phi\log\tau\,,
\end{equation}
where $c_t$ and $c_x$ are positive numbers and
\begin{equation}
p_t=\frac{\mathsf{z}-2}{\mathsf{z}+2-2\theta},\qquad p_x=\frac{2-\theta}{\mathsf{z}+2-2\theta},\qquad p_\phi=\frac{\sqrt{2(2-\theta)(2\mathsf{z}-2-\theta)}}{\mathsf{z}+2-2\theta}.
\end{equation}
The Kasner conditions $p_t+2p_x=1$ and $p_t^2+2p_x^2+p_\phi^2=1$ are satisfied.

\section{AdS$_5$ solutions}
\label{sec:AdS5}
We briefly summarize the AdS$_5$ counterpart of section~\ref{sec:GS}. We take the potential of the scalar field $\phi$ given by \cite{Gao:2004tv}
\begin{equation}
V(\phi) =-\frac{12}{(4+3\alpha^2)^2L^2}\Bigl[3\alpha^2(3\alpha^2-2)e^{-\frac{4\phi}{3\alpha}}
+36\alpha^2e^{\frac{3\alpha^2-4}{6\alpha}\phi}+2(8-3\alpha^2)e^{\alpha\phi}\Bigr],
\label{eq:V-AdS5}
\end{equation}
where $\alpha$ is a parameter, and the values of $\alpha=0$, $2/\sqrt{6}$ and $4/\sqrt{6}$ correspond to special cases of STU supergravity. The values of $\alpha=2/\sqrt{6}$ and $4/\sqrt{6}$ correspond to 2-charge and 1-charge black holes in AdS$_5$, respectively.

The $\phi\to 0$ behavior is $V(\phi)=-12/L^2-(2/L^2)\phi^2+\cdots$, where the first term is the cosmological constant, and the second term gives the mass of the scalar field $m^2L^2=-4$. The scaling dimension of the dual scalar operator in the CFT is $\Delta_\pm=2$. Unlike the AdS$_4$ case, we can use the standard quantization without deformation, and the alternative quantization is not available.

\begin{figure}
\centering
  \includegraphics[]{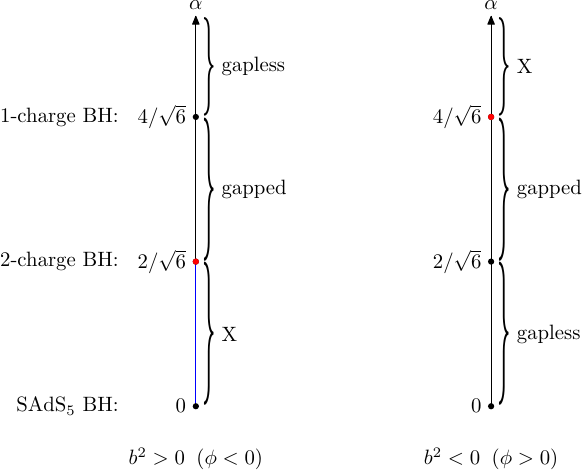}
  \caption{\label{fig:IR-AdS5} Classification of the one-parameter family of neutral solutions in AdS$_5$ according to their IR geometries. Special cases belonging to supergravity are marked. SAdS$_5$ denotes the planar Schwarzschild-AdS$_5$ black hole. ``X'' denotes that in this range of parameter space, the hyperscaling-violating geometry violates the Gubser criterion. In the blue interval ($0\leq\alpha\leq 2/\sqrt{6}$), a class of charged black hole solutions exists, and the IR geometries are changed by the gauge field.}
\end{figure}

The system admits a neutral solution with a nontrivial dilaton profile. The solution of the metric $g_{\mu\nu}$, gauge field $A_\mu$, and dilaton field $\phi$ is \cite{Ren:2019lgw}
\begin{align}
ds^2 &=f\tsp (-dt^2+d\vec{x}^2)+g^{-1}dr^2,\qquad A=0\,,\nonumber\\
f &=\frac{r^2}{L^2}\left(1-\frac{b^2}{r^2}\right)^\frac{3\alpha^2}{4+3\alpha^2},
\qquad 
g=\frac{r^2}{L^2}\left(1-\frac{b^2}{r^2}\right)^{\frac{6\alpha^2}{4+3\alpha^2}},\qquad
e^{\alpha\phi} =\left(1-\frac{b^2}{r^2}\right)^\frac{6\alpha^2}{4+3\alpha^2},
\label{eq:solution5}
\end{align}
where the parameter $b^2$ can be either positive or negative. For $\alpha\neq 0$, this metric does not have a regular horizon, and has a spacetime singularity at $r=b$ if $b^2>0$ ($b>0$), and at $r=0$ if $b^2<0$. Classification by IR geometries is summarized in figure~\ref{fig:IR-AdS5}. We expect that this solution can be taken as an extremal limit of a finite temperature black hole solution, when the Gubser criterion is satisfied.

\end{document}